# An Agent-Based Intelligent HCI Information System in Mixed Reality

**Hamed Alqahtani**  *hsqahtani@kku.edu.sa*
*VISOR, Faculty of Science and Engineering, Macquarie University, Sydney, Australia*

**Charles Z. Liu**  *charles.liu@mq.edu.au*
*VISOR, Faculty of Science and Engineering, Macquaire University, Sydney, Australia*

**Manolya Kavakli-Thorne**  *manolya.kavakli@mq.edu.au*
*VISOR, Faculty of Science and Engineering, Macquarie University, Sydney, Australia*

**Yuzhi Kang**  *y.kang@psismartsys.com*
*PSI Smartsys, Sydney, Australia*

### Abstract

This paper presents a design of agent-based intelligent HCI (iHCI) system using collaborative information for MR to improve user experience and information security based on context-aware computing. In order to implement target awareness system, we propose the use of non-parameter stochastic adaptive learning and a kernel learning strategy for improving the adaptivity of the recognition. The proposed design involves the use of a context-aware computing strategy to recognize patterns for simulating human awareness and processing of stereo pattern analysis. It provides a flexible customization method for scene creation and manipulation. It also enables several types of awareness related to the interactive target, user-experience, system performance, confidentiality, and agent identification by applying several strategies, such as context pattern analysis, scalable learning, data-aware confidential computing.

## 1. Introduction

Virtual Reality (VR) deals with an alternate world filled with simulated environments and objects. Computer-generated images allow the end user to interact with the virtual space. The virtual reality presents the user the experience of a mixture of reality and virtuality [6, 12, 54]. Interaction in VR can be divided into three types; namely, textual VR refers to interaction without visualization; desktop VR deals with interaction with visualization, and immersive VR handles interaction with a high level of immersion [4]. The immersive interaction has serious issues in VR [3, 47]. Augmented Reality (AR) refers to an experience in a direct or indirect view of a physical, real-world environment. AR systems are more widely used than immersive VR systems due to their low cost and easily accessible characteristics [41, 22, 17, 60]. However, a mixture of virtual and real worlds supported with equally physical and augmented elements requires immersive interaction.

Thereafter, Mixed reality (MR) systems [39, 55, 7] were proposed to perform merging of real, with a cyber world, providing possible improvements for high-level of interaction, such as memory functioning [43], attention enhancement [5] and behaviour therapy [19]. Modern computer systems facilitated with auxiliary functionalities are capable of performing analysis on data and visualization with marked characteristics [37, 2]. However, these devices require some sensors and actuators (as in [13]) for such interaction. The interactions developed still lack the understanding of users and their intention.

This paper presents the development of an intelligent HCI (iHCI) system using a piece of collaborative information for MR [32, 25] and allows the system to understand the user's intention during the interaction. Such interaction requires the systems to be aware of the subject and context in human-computer interaction (HCI). The proposed system enables a better user experience and information security using context-aware computing [16, 10].

Rest of the paper is organized as follows. Section 2 provides a review of state of the art in the field of MR. Section 3 describes the agent-aware computing. Section 4 describes the system and detailed design of the proposed system. Section 5 presents the implementation details of the proposed system. Finally, Section 6 concludes the paper at the end.

## 2. Related Work

The applications of computational intelligence in virtual environments are a convergence of smart system and HCI systems. Development of intelligent HCI (iHCI) systems remained as a key to this convergence. Several designs have been proposed in the last decade.

The agent-based design is one of the most commonly used designs for implementing context-aware iHCI systems. Several designs have been proposed using agents for developing intelligent system design [24, 50, 18, 35, 48]. These designs have different capabilities for intelligence, such as perception, cognition, recognition and instruction. However, there are many issues related to intelligent VR systems that include knowledge representation, autonomous agents, and agent coupling. Those intelligent interactions were based on the results of machine states influenced by the users' operation.

Most of the network-based applications mainly have focused on communication, traffic flow and implementation of the interactive systems [45, 21, 9]. These applications of computational intelligence have involved interactive elements, such as virtual objects, virtual roles and real user connected by the network. The issue of optimization of traffic flow was also addressed in those works. However, iHCI systems mostly focused on the performance of the system, rather than the performance of user experience.

Many studies have been conducted to explore the effects on users during the interaction [34, 20, 40]. The factors that affect the users include graphic displays, dialogue design, and operation interface, motivate for improving the user-experience in HCI. However, an iHCI system must perform a real-time user experience assessment. It is a challenge for a machine to understand abstract concepts like user experience.

Knowledge systems can help HCI systems to acquire a human sense of interaction and experience. Many researchers focused on knowledge translation and knowledge system implementation for system operations [51, 1, 23, 44]. The knowledge in these systems is mainly used as a logic of rules and data. However, an iHCI system must take the human perception and cognition into account to simulate the real response by a human rather than a logic link between the rules and the knowledge base. Human perception is abstract possessing essential characteristics of uncertainty. Subjective awareness is hard to quantify, while computation needs to be specific and accurate for the processing of interactive data. These issues are considered as a challenge for an intelligent HCI system.

Moreover, the systems must be aware of its own operations for considering wide area mixed reality applications running in an open field of interaction [36, 57]. Because of information about users' behaviour and response that will be generated during immersive interaction. This type of information may be sensitive as well, in terms of privacy, especially in therapy, education, and military [8, 38, 49]. If the system had the self-awareness to detect such anomalousness, it can detect the unusual changes and protect the system as well as users' privacy. Therefore, awareness enables the HCI systems to perform an advanced interaction with humanized experience and information security.

In order to perform such awareness, a system may need sensors to collect information about the users and itself. This design implies that the system is intricate with considering the amount of complexity required for processing. Consequently, the cost of establishing the cyber environment can be slashed due to the advantages brought by the sensors.

These issues motivate us to develop a mixed reality system for enhanced awareness. So, we explore solutions to enable an HCI system to perform smart interaction with agent aware computing enhanced intelligence and achieve a vision-based scheme of the mixed reality system. Thus, it provides support for the implementations of vision based smart interactive systems for mixed reality applications.

## 3. Agent-Aware Computing

The concept of Agent Aware Computing (AAC) refers to computational processing for enabling the machine to be aware of an agent operating in the system. This may include humans, devices and the combination of the humans and devices. Such systems can analyze the features, and attributes of an agent using an awareness model. The targets and subjects involved in the interaction with the data captured as the result of the analysis will be stored as knowledge for processing to complete a certain task. In terms of enhancing the interaction in a mixed reality system (i.e., enhanced mixed reality, eMR), we divide the tasks of agent aware computing into three main groups: human aware, system aware, and human- system aware.

Human-aware computing (HAC) targets to simulate human visual attention, and memory. The main tasks of HAC include simulating concentration to the target a) to distinguish the goal in the scene and b) to differentiate it from the other object as a background, and memory fusion, i.e., to fuse the target of interest with another scene;

System-aware computing (SAC) targets to make the system aware of its devices, network, and operations related to performance and processing. The main task of SAC includes a) performing the authentication to the tools to detect the well-intended and malicious activities; b) being aware of the structure and activities of the system for the processing, such as knowledge learning and updating data;

Human-system-aware computing (HSAC) targets to make the system aware of both sides of human and system in processing. The main tasks of HSAC are to perform the comprehension of both the user experience and system performance. The methodological functions to support the agent-aware computing scheme, include: 1) target of interest (ToI) - aware computing: to make the system aware of the subject of interaction as a target of interest; 2) context-aware computing: to make the system aware of the set of circumstances or facts during the interaction and processing; 3) user-experience aware computing: to enable the system to be aware of the experience of a user during the interaction. The context in context-aware computing consists of two orientations: 1) context of interaction: the context of the scenario related to the perception of real world that surrounds the subject in a particular event, situation, etc.; 2) context of the system environment: the context of the cyber environment related to the devices, operations, and processes to perform the functionalities of the system.

## 4. The proposed design

This section provides the design of the proposed system by explaining system design and detailed design with the interaction of different components.

### 4.1. System Design

Human vision is approximately ten times as much as the depth detector. In some cases, the texture of the subjects is more important than the distance, for example, in applications (such as [52, 46, 14, 42, 56]). With visual computing, the system's dependence on the sensors is removed. This brings more flexibility to the implementation. For example, participants would no longer need to wear a suit with many sensors and transmitters to communicate with the receivers in a virtual environment. Responses of the system are delivered according to the visual information captured by other cameras or other types of optical sensing technologies (e.g., infrared sensors). Therefore, the interaction would be simplified in terms of the calibration for networking as well as transmitting-receiving test. Meanwhile, the number of participants will not be limited by the number of devices or suit. Special effects can be added to the simulation, which brings users more vivid experience during the interaction. Thus, it provides another option for acquiring better visual effect with less device dependence.

There are several issues in implementing a collaborative information system for MR [29]. The major issues include scenario awareness and target awareness, knowledge and experience modelling, scalable knowledge management, confidentiality and security. The design of the system is divided into three main subsystems: Mixed Reality Fusion Subsystem (MRFS), User Awareness Subsystem (UAS) and Sensing Network Subsystem (SNS), as shown in Figure 1. Each subsystem is supported by related modules.

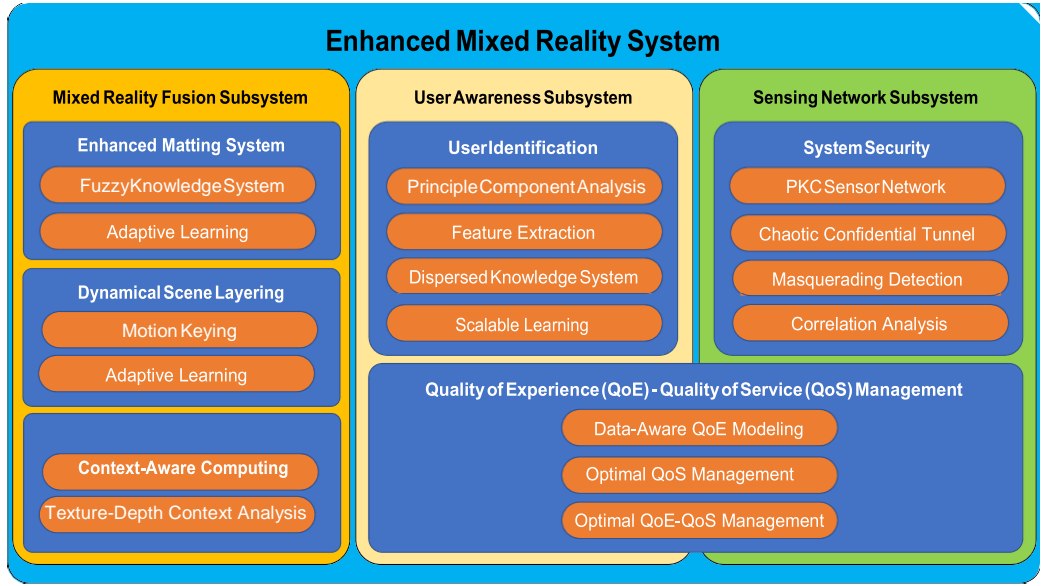

Figure 1: The Modules of the Enhanced Mixed Reality System

In the MRFS subsystem, Enhanced Matting System (EMS), Dynamical Scene Layering (DSL), and Context-Aware Computing (CAC) support the system to adapt to the texture, motion and depth data; In the UAS subsystem, User Identification Module (UIM) supports the system to recognize the user, providing the corresponding information for interaction; In the SNS subsystem, System Security Module (SSM) is designed for the information system to guarantee the information security in collection, organization, storage and communication of information during the HCI interaction.

Quality of Experience (QoE) - Quality of Service (QoS) Management (QoE- QoS-M) module serves as a joint module for both the UAS and SNS. QoE and QoS play an important role in both user awareness and the network system. With the support of the QoE-QoS-M, the balance between user experience and system performance can be used to improve the usability of the whole system.

In this paper, texture-based recognition, motion-based recognition and context-aware computing, are studied and implemented to enhance the differentiation. Principles of optimization to reduce the uncertainties are studied and discussed. Corresponding optimization strategies are proposed to guarantee the stability of the solution.

QoE-QoS management is applied to assess the quality of the balance between user experience and communication performance. This module is mainly responsible for the awareness of user experience and system performance. We implemented a data-aware strategy for QoE-QoS management. Three main types of QoE-QoS management (optimal QoE management, optimal QoS management, and QoE-QoS balance management) are discussed. With the consideration of user-experience and system performance, a unified optimization strategy with quantification and normalization of data space has been proposed to perform QoE- QoS balancing.

After the assessment, the data volume is re-encoded in a compression form. A protocol with PKC based cryptography is designed to perform the encryption and decryption to guarantee the confidentiality of the data flow and authentication, implementing the awareness of privacy and agent identification (in terms of machinery) to reinforce the security of the system. With the consideration of both efficiency and performance, a protocol has been implemented to establish a chaotic tunnel for logical peer to peer communication. The system can detect anomalous node in the network and protect both the information and communication against attacks. Related strategies are designed with the consideration of the critical space, sensitivity, correlation and uniform distribution to enhance the self-awareness of the system.

### 4.2. System Architecture

Figure 2 shows the system architecture and the workflow among the components. Vision is used as the information source to overcome the constraints of a sensor-based data system. In the proposed system as depicted in Figure 2, the input is from imaging systems that provide visual data of the target scene. Since high-resolution data will be expensive in terms of computational resource, before transmission, the data is processed by QoE-QoS management system. This brings a balance between user experience and communication performance and provides the awareness of user experience and system performance.

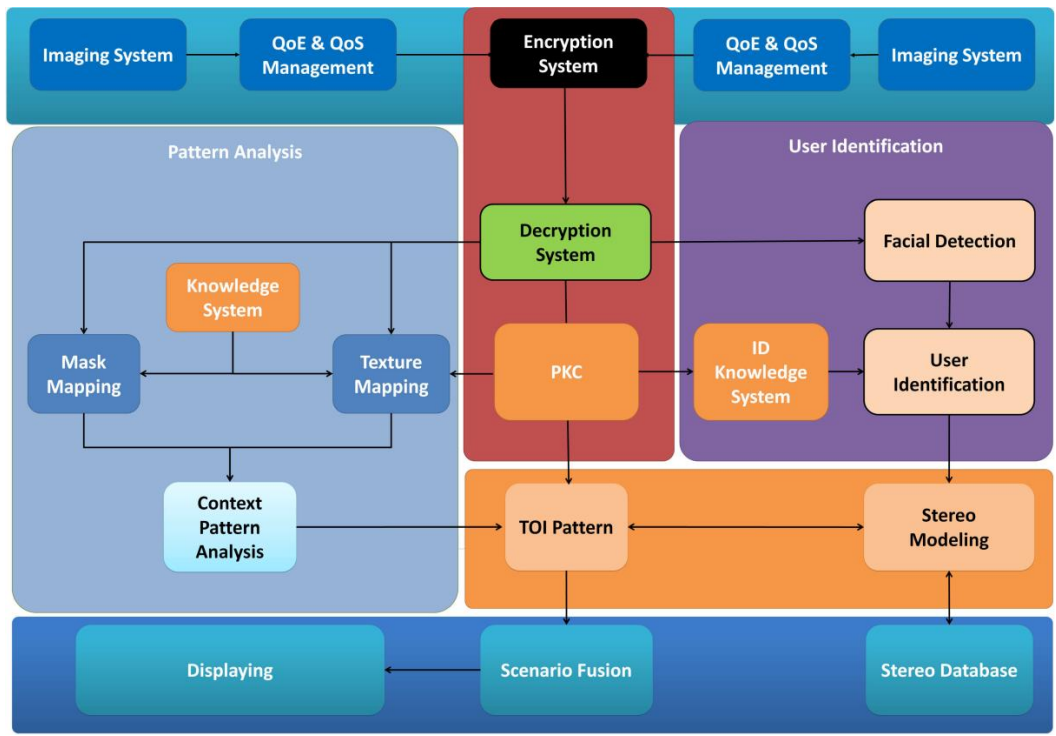

Figure 2: System Synthesis and Workflow

Considering information security, we introduced a protocol with PKC based cryptography to perform encryption and decryption to guarantee the confidentiality of the dataflow. In this part, the awareness of privacy and agent identification (in terms of machinery) is implemented to reinforce the security of the system. The authorized data moves to pattern analysis to differentiate the target of interest (TOI) in the captured scene. This supports stereo modelling and scenario fusion. For privacy management, a user identification module is developed. This implements the agent identification in the view of the human.

### 5. Implementation

In order to simulate human visual attention and memory, we propose to use visual computing as visual computing is closer to the way how humans perceive. It also alleviates the constraints on the implementation, which suits the goals of aware human computing well. We apply recursive learning and fuzzy adaptive learning strategies to form awareness gradually [27]. The target of interest recognition is performed by a recursive α optimization. Corresponding knowledge is stored as a fuzzy set with learning parameters.

During the implementation, modules are designed according to the works [29, 27, 28, 26, 30, 31], and corresponding processing is synthesized as an integrated system. The system is designed under a CIS architecture (as shown in Figure 3), where the agent in the system can be either a human user or machinery (computer, camera, or other devices) or combination (user with a machine). All agents can collect data from and upload to the system depending on their roles.

To implement target awareness, we apply a non-parameter stochastic adaptive learning strategy [33]. Based on Gaussian Modeling (GM), a kernel learning strategy is designed to improve the adaptivity of the recognition. This enables the system to differentiate the target by motion cues and perform a scenario layering task without the support of prior knowledge of the scene. A context-aware computing strategy to recognize patterns is designed and applied to simulate human awareness and processing of stereo pattern analysis [28]. Based on the knowledge-based inference system, human experience can be used to model the correlations of the patterns in the context. We have developed an enhanced learning approach to allow the system to process the ambiguous patterns and to refine confidence.

To implement system-awareness (SA), we apply data-aware computing using authentication. With the security enhancement for confidential communication, the system can be aware of the authorities of its devices, in the network and detect the potential malicious attacks, using public key cryptography [31]. This brings system awareness in terms of system security. Regarding operations, we focus on learning and knowledge management to contribute to system-awareness. System-awareness has been modelled as a process [30]. A scalable learning strategy is designed to simulate the process of knowledge broadening and applied to perform scalable management using a dispersed structure.

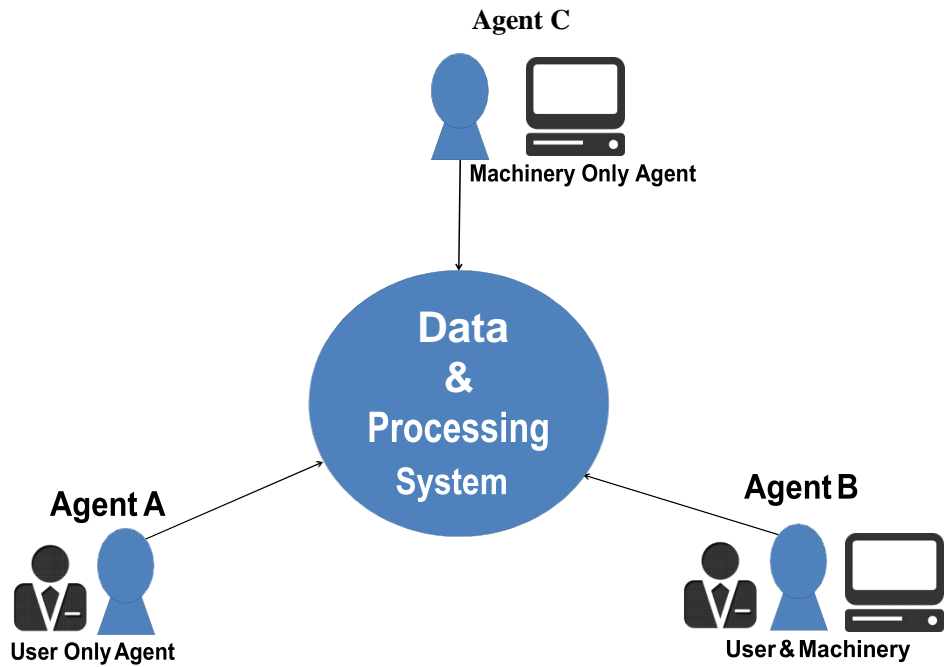

Figure 3: The Scheme of The Collaborative Information System

To implement human-system-aware computing (HSAC), we model the comprehension of user experience and system performance based on subspace learning with the quality of experience (QoE) and the quality of service (QoS) and use the model to estimate user experience [26], by sensing the service. A camera system is designed for capturing the visual data. For performing stereo modelling, a multiple-view camera system is applied, in which two cameras capture the visual information from the user's right-hand side profile and frontal view. We project the user's image into a virtual reality platform.

Using data-aware user experience modelling [26] to compress the data, we improve the efficiency of communication. MOS (mean of score) is used to measure the experience. A data-aware confidential communication strategy is used in the system. The information of the users has been ciphered, and the information fingerprint has been given by using HASH. The data fingerprint by HASH is used to implement awareness of agent identification. By matching the finger-print, the data and its source can be detected, as well as the authorization of the components linked to the system.

The captured data is ciphered with applying a data-aware confidential PKC strategy [31]. For mixed reality fusion (MRF), the system must recognize the user as a target of interest (TOI) firstly and then map the user into virtual and mixed reality, and users' facial information is stored in the system database for identification. A dispersed database is used to store the data of different users. Since the data is distributed, the information can be accessed and updated quickly, and scalable learning strategy [30] is used to train the knowledge system for identification.

After TOI processing, the data of a user and an outdoor scene are transferred to process the scene fusion (as shown in Figure 3). Based on the results above, a recognition based mixed reality fusion using a real-virtual-object (RVO) has been further implemented. This function provides a flexible customization method for scene creation and manipulation. The targeted scene is selected as a background for the scene fusion. By pattern-context based analysis and optimal context fusion [28, 27], the user's image is synthesized with the target scene, and the result is transferred to display.

## 6. Conclusion

This paper presents a design of agent-based intelligent HCI (iHCI) system using a piece of collaborative information for MR to improve user experience and information security on the basis of context-aware computing. Several types of awareness related to the interactive target, user-experience, system performance, confidentiality, and agent identification, have been implemented by applying a number of strategies, such as context pattern analysis, scalable learning, data-aware confidential computing, in the implementation. The proposed design enables the addressing to some of the significant issues like mixed reality fusion, QoE-QoS management, confidentiality, and security. Applying agent-aware computing-based scheme into an implementation of the collaborative information system for mixed reality, we improve user experience and immersion. The proposed system using agent-aware computing scheme provides a valid approach to tackle the issues, regarding the lack of understanding of users.